\documentclass[a4paper,11pt]{article}
\pdfoutput=1 % if your are submitting a pdflatex (i.e. if you have images in pdf, png or jpg format)
\usepackage{jcappub} % for details on the use of the package, please see the JCAP-author-manual
\usepackage[T1]{fontenc} % if needed

\title{\huge Hyperbolic field space and swampland conjecture for DBI scalar}

\author[a]{\large Shuntaro Mizuno,}
\author[b,c]{\large Shinji Mukohyama,}
\author[c]{\large Shi Pi,}
\author[b]{\large Yun-Long Zhang}

\affiliation[a]{Department of Liberal Arts and Engineering Sciences, National Institute of Technology, Hachinohe College, Aomori, 039-1192, Japan}
\affiliation[b]{Center for Gravitational Physics, Yukawa Institute for Theoretical Physics,\\ Kyoto University, Kyoto 606-8502, Japan}
\affiliation[c]{Kavli Institute for the Physics and Mathematics of the Universe (WPI), \\The University of Tokyo Institutes for Advanced Study,\\The University of Tokyo, Kashiwa, Chiba 277-8583, Japan}

\emailAdd{mizuno-g@hachinohe.kosen-ac.jp}%shuntaro.mizuno@yukawa.kyoto-u.ac.jp}
\emailAdd{shinji.mukohyama@yukawa.kyoto-u.ac.jp}
\emailAdd{shi.pi@ipmu.jp}
\emailAdd{yun-long.zhang@yukawa.kyoto-u.ac.jp}

\abstract{We study a model of two scalar fields with a hyperbolic field space and show that it reduces to a single-field Dirac-Born-Infeld (DBI) model in the limit where the field space becomes infinitely curved. We apply the de Sitter swampland conjecture to the two-field model and take the same limit. It is shown that in the limit, all quantities appearing in the swampland conjecture remain well-defined within the single-field DBI model. Based on a consistency argument, we then speculate that the condition derived in this way can be considered as the de Sitter swampland conjecture for a DBI scalar field by its own. The condition differs from those proposed in the literature and only the one in the present paper passes the consistency argument. As a byproduct, we also point out that one of the inequalities in the swampland conjecture for a multi-field model with linear kinetic terms should involve the lowest mass squared for scalar perturbations and that this quantity can be significantly different from the lowest eigenvalue of the Hessian of the potential in the local orthonormal frame if the field space is highly curved. Finally, we propose an extension of the de Sitter swampland conjecture to a more general scalar field with the Lagrangian of the form $P(X,\varphi)$, where $X=-(\partial\varphi)^2/2$.}

\keywords{Swampland Conjecture, DBI scalar, Two-Field Model}
\arxivnumber{1905.10950}
\proceeding{{\small YITP-19-42, IPMU19-0082} 
}
\dedicated{
{\bf DOI}:
${\href{https://doi.org/10.1088/1475-7516/2019/09/072}{https://doi.org/10.1088/1475-7516/2019/09/072}}$\hfill
Nov. 22, 2019}

\begin{document}
\maketitle
\flushbottom
 
\section{Introduction}

The universe today is expanding at an accelerated rate~\cite{Riess:1998cb,Perlmutter:1998np},  which in the context of general relativity suggests the existence of a mysterious component dubbed ``dark energy''. In the very early universe, the most well studied mechanism that can set up the initial condition of the hot big bang is inflation, i.e. another accelerated expansion~\cite{Brout:1977ix,Sato:1980yn,Guth:1980zm,Linde:1981mu,Albrecht:1982wi,Starobinsky:1979ty,Mukhanov:1981xt}. From the effective field theory point of view, both of these accelerated expansions can be described by one or some scalar fields which move slowly in the plateau-like potential with some positive energy density. From the viewpoint of quantum gravity, on the other hand, to realize such a quasi-de-Sitter spacetime is not a trivial task. For instance, a certain no-go theorem against $4$-dimensional solutions with accelerated expansion and stabilized moduli have been known for a class of gravitational theories~\cite{Maldacena:2000mw}. Extending the no-go theorem,  Obied \textit{et al.}~\cite{Obied:2018sgi} recently conjectured that for a theory coupled to gravity with the potential $V$ of scalar fields, a necessary condition for the existence of a Ultraviolet (UV) completion is
\begin{equation}
\left|\nabla V\right|\geq c\,V,
\end{equation}
which is called the de Sitter conjecture \footnote{Although the authors of \cite{Obied:2018sgi} named this
de Sitter conjecture, since this conjecture belongs to the wider ones named swampland conjectures, it is often called de Sitter swampland conjecture. Therefore, in this paper, we will also call this as de Sitter swampland conjecture from now on. For the earlier discussion of the swampland conjectures, see \cite{Ooguri:2006in, Brennan:2017rbf}}. In the above inequality, the norm $\left|\nabla V\right|$ of the gradient of the potential in the field space is defined using the field space metric that can be read off from the kinetic term of the scalar fields, $c$ is a positive constant of $\mathcal{O}(1)$, and the Planck mass $M_{\rm Pl}$ is set to unity. 
However, since this was just the conjecture, there was some drawback like the existence of
dS extrema \cite{Roupec:2018mbn} and supersymmetric AdS vacua \cite{Conlon:2018eyr}.
Also, some people like the authors of \cite{Kachru:2018aqn,Akrami:2018ylq,Murayama:2018lie} 
took the skeptical attitude to the conjecture or tried to improve it. 
With this situation, more recently, Ooguri \textit{et al.}~\cite{Ooguri:2018wrx} modified this conjecture to what is called the refined de Sitter swampland conjecture, i.e. 
\begin{equation}
\left|\nabla V\right|\geq c\,V,  \quad \mbox{or}\quad \text{min}\left(\nabla_I\nabla_J V\right)\leq-c'\,V\,, \label{eqn:conjecture-slowroll}
\end{equation}
where $c'$ is another $\mathcal{O}(1)$ positive constant, and $\text{min}\left(\nabla_I\nabla_J V\right)$ is the minimum eigenvalue of the Hessian of the potential in the local orthonormal frame. Here we use $I, J$ to denote the indexes of the scalar fields. For more recent
discussions on the improvement and extension of this conjecture, 
see e.g.~\cite{Dasgupta:2018rtp,Danielsson:2018qpa,Motaharfar:2018zyb,Hebecker:2018vxz,Dvali:2018jhn,Junghans:2018gdb,Andriot:2018mav,Kallosh:2019axr,Blumenhagen:2019qcg,Font:2019cxq,Heckman:2019bzm,Garg:2018zdg,Das:2018rpg}
and \cite{Palti:2019pca}  for a review.

This conjecture has been widely applied to cosmology like inflation and dark energy.
For inflation, there are discussions on the inflation of single field ~\cite{Garg:2018reu,Dias:2018ngv,Kinney:2018nny,Lin:2018kjm,Brahma:2018hrd,Das:2018hqy,Kawasaki:2018daf,Ashoorioon:2018sqb,Fukuda:2018haz,Park:2018fuj,Lin:2018rnx,Yi:2018dhl,Holman:2018inr,Artymowski:2019vfy,Haque:2019prw,Sabir:2019wel,Benetti:2019smr,Ben-Dayan:2018mhe,Scalisi:2018eaz,Dimopoulos:2018upl}, 
of multi-field ~\cite{Bjorkmo:2019aev,Achucarro:2018vey,Schimmrigk:2018gch,Fumagalli:2019noh,Lynker:2019joa,Micu:2019fju,Damian:2018tlf,Aragam:2019khr,Cheong:2018udx,Bjorkmo:2019fls}, 
and with non-canonical kinetic term~\cite{Kehagias:2018uem,Seo:2018abc,Bhattacharya:2018xlw,Herdeiro:2018hfp,Heisenberg:2019qxz}. For dark energy, it is clearly seen that a cosmological constant is not compatible with the swampland conjecture, while a dynamical quintessence field is favored~\cite{Agrawal:2018own,Dvali:2018fqu,Chiang:2018jdg,DAmico:2018mnx,Odintsov:2018zai,Olguin-Tejo:2018pfq,Heckman:2018mxl,Chiang:2018lqx,Russo:2018akp,Cai:2018ebs,Heckman:2019dsj,Nan:2019gjw,Almeida:2019iqp,vandeBruck:2019vzd,Han:2018yrk,Denef:2018etk,Ibe:2018ffn,Wang:2018kly,Cicoli:2018kdo,Marsh:2018kub,Emelin:2018igk,Hertzberg:2018suv,Matsui:2018xwa,Acharya:2018deu}. The  constants $c$ and $c'$ can be constrained by the current and future observational constraints on the evolution of equation of state $w_\text{DE}$ for the dark energy~\cite{Andriot:2018wzk,Colgain:2018wgk,Heisenberg:2018yae,Heisenberg:2018rdu,Wang:2018duq,Agrawal:2018rcg,Elizalde:2018dvw,Tosone:2018qei,Raveri:2018ddi,Brahma:2019kch,Colgain:2019joh}.

The (refined) de Sitter swampland conjecture was originally formulated for scalar fields with linear kinetic terms, from which one can read off the field space metric. On the other hand, string theory allows for not only scalar fields with linear kinetic terms but also Dirac-Born-Infeld (DBI) scalar fields with nonlinear kinetic terms~\cite{Silverstein:2003hf,Alishahiha:2004eh}. It is therefore natural to ask whether one can extend the conjecture to DBI scalars or not. A DBI scalar field is a special case of a k-essence type scalar field~\cite{ArmendarizPicon:1999rj,Garriga:1999vw,ArmendarizPicon:2000ah}. Let us therefore consider a general k-essence type scalar field with the Lagrangian of the form $P(X, \varphi)$, where $X=-g^{\mu\nu}\partial_{\mu}\varphi\partial_{\nu}\varphi/2$ and $\mu, \nu = 0,\cdots,3$ denote the spacetime indexes. Apparently, there seem at least three possibilities for extension of the de Sitter swampland conjecture to such a scalar field. The option (i) is to expand the Lagrangian around $X=0$ as
\begin{equation}
 P(X, \varphi) = P_0(\varphi) + P_1(\varphi)X + \mathcal{O}(X^2)\,,
\end{equation}
and to make the following identification
\begin{equation}
 V(\phi) \ \Leftrightarrow \ -P_0(\varphi)\,, \quad 
 d\phi \  \Leftrightarrow \ \sqrt{P_1(\varphi)}d\varphi
\end{equation}
in (\ref{eqn:conjecture-slowroll}). This option is justified in the slow-roll limit but does not have a clear justification beyond the slow-roll regime. The option (ii) is to consider a linear perturbation around a homogeneous background as $\varphi = \varphi^{(0)}(t)+\pi(t,\vec{x})$, where $t=x^0$ is the time coordinate and $\vec{x} = (x^1, x^2, x^3)$ are spatial coordinates, and to read off an alternative to the field space metric from the (time) kinetic term for the perturbation. Expanding the quadratic action for $\pi$ as
\begin{equation}
 P(X,\varphi)\ \ni \ \frac{1}{2}\mathcal{K}_{\parallel}\dot{\pi}^2 - \frac{1}{2a^2}\mathcal{K}_{\perp}\delta^{ij}\partial_i\pi\partial_j\pi\,,
\end{equation}
where $\mathcal{K}_{\parallel} = (2P_{,XX}X+P_{,X})^{(0)}$, $\mathcal{K}_{\perp} = P_{,X}^{(0)}$, $\dot{\pi}\equiv\partial_t \pi$ and $a$ is the scale factor of a Friedmann-Lemaitre-Robertson-Walker (FLRW) spacetime.
We can make the following identification
\begin{equation}
 d\phi \  \Leftrightarrow \ \sqrt{\mathcal{K}_{\parallel}}d\varphi\,, 
\end{equation}
supposing that the no-ghost condition $\mathcal{K}_{\parallel}>0$ is satisfied. This option uses the (time) kinetic term for the perturbation to define an alternative to the field space metric but does not provide a replacement for the potential $V(\phi)$. The option (iii) is to make the following identification
\begin{equation}
 d\phi \  \Leftrightarrow \ \sqrt{\mathcal{K}_{\perp}}d\varphi\,, 
\end{equation}
supposing that the no-ghost and no-gradient-instability conditions $\mathcal{K}_{\parallel}>0$, $\mathcal{K}_{\perp}>0$ are satisfied. This option uses the gradient term for the perturbation to define an alternative to the field space metric but again does not provide a replacement for the potential $V(\phi)$. Unfortunately, none of the three arguments is convincing.

One of the main purposes of the present paper is to propose a more promising option for the extension of the swampland conjecture to a DBI scalar field and a general k-essence type scalar field. Our proposal is based on the observation that a certain model of two scalar fields with a curved field space is reduced to a single-field DBI or k-essence scalar model in the limit where one of the fields becomes infinitely heavy\footnote{Although we do not consider in this paper, it is worth mentioning that making use of a curved field space, interesting cosmological scenarios like geometrical destabilization \cite{Renaux-Petel:2015mga} and hyperinflation \cite{Brown:2017osf,Mizuno:2017idt} have been proposed recently.
See also \cite{Anguelova:2018vyr} on the recent discussion of the symmetry.}. By applying the refined de Sitter swampland conjecture to the two-field model and then taking the limit, we find the extension of the conjecture to a DBI scalar field and a general k-essence type scalar field.

The rest of the present paper is organized as follows. In section \ref{secSW}, we review the refined swampland conjecture for scalar fields with linear kinetic terms. In section \ref{secHS}, we study the attractor behavior for the two-field model in the hyperbolic field space and its squared masses of scalar perturbation modes. In section \ref{secDBI}, we study the swampland conjecture for the DBI scalar, which can be reduced from the hyperbolic field space. In section \ref{secPX}, we follow the logic and propose the swampland conjecture for more general $P(X,\varphi)$ theories.
We make the summary and discussion in section \ref{secCon}.

\section{Refined Swampland conjecture with linear kinetic terms\label{secSW}}

We now briefly review the refined de Sitter swampland conjecture \cite{Ooguri:2018wrx} for scalar fields with linear kinetic terms. In the following, we work in the Einstein frame and adopt the unit with $M_{\rm Pl}=1$.

The distance conjecture~\cite{Ooguri:2006in} is originally a statement about the moduli space of the string landscape. It states that (i) the moduli space is parametrized by expectation values of scalar fields $\{{\phi}^I|I=1,2,3\dots\}$ with linear kinetic terms of the form
\begin{equation} 
 L_{\rm kin} = -\frac{1}{2}\mathcal{G}_{IJ}(\phi^K)\, g^{\mu\nu}\partial_{\mu}\phi^I\partial_{\nu}\phi^J\,, \label{eqn:Lkin}
\end{equation}
and a flat potential $V({\phi}^I)=0$, that (ii) the moduli space includes points with infinite geodesic distances from each other and that (iii) towers of light states with masses of order $m\sim e^{-a\Delta\phi}$ appear as we move the geodesic distance $\Delta\phi$ ($\gg 1$) away from a point in the moduli space. Here, $a$ is a positive number of $\mathcal{O}(1)$ and the geodesic distance in the moduli space is defined by the metric $\mathcal{G}_{IJ}({\phi}^K)$ in (\ref{eqn:Lkin}), which is a function of the scalar fields $\{{\phi}^K\}$. 

One of the basic elements in the argument of ref.~\cite{Ooguri:2018wrx} is the assumption that the distance conjecture holds not only in the moduli space with a flat potential but also in a field space with the kinetic lagrangian $L_{\rm kin}$ and a non-trivial potential $V({\phi}^I)\ne 0$. It then suggests that the number of particle species $N$ below the cutoff of an effective field theory is roughly given by 
\begin{equation}
 N \sim n(\phi)e^{b\phi}\,, \quad \frac{dn}{d\phi} > 0\,, \label{eqn:N-ansatz}
\end{equation}
where $\phi$ is the geodesic distance from a point in the field space deep inside the regime of validity of the effective field theory, $n(\phi)$ represents the effective number of towers of light states and $b$ is a positive number of $\mathcal{O}(1)$. Namely, each tower has an exponentially large number of light particles and there are $n(\phi)$ towers. Another element in the argument is the following ansatz for the entropy of the towers of light particles in an accelerating universe
\begin{equation}
 S_{\rm tower}(N,R) \sim N^{\delta_1} R^{\delta_2}\,, \label{eqn:entropy-ansatz}
\end{equation}
where $N$ is the number of particle species, $R\sim 1/H$ is the radius of the apparent horizon, $H$ is the Hubble expansion rate and $\delta_{1,2}$ are positive numbers of $\mathcal{O}(1)$.

Yet another basic element is the covariant entropy bound~\cite{Bousso:1999xy}, conservatively applied to a quasi de Sitter spacetime in the Einstein frame. While the covariant entropy bound is considered to be applicable to a wider class of FLRW spacetimes, being conservative, ref.~\cite{Ooguri:2018wrx} considered the following condition as a sufficient condition for the applicability of the covariant entropy bound. 
\begin{equation}
 \left|\frac{\dot{H}}{H^2}\right| \lesssim c_1\,, \quad
  \frac{\min m_{\rm scalar}^2}{H^2} \gtrsim -c_2\,, \label{eqn:condition-for-entropy-bound}
\end{equation}
where an overdot represents derivative with respect to the proper time, $\min m_{\rm scalar}^2$ is the lowest among squared masses of perturbation modes of the scalar fields, and $c_{1,2}$ are positive numbers of $\mathcal{O}(1)$. The first inequality states that the geometry is close to a de Sitter spacetime and the second states that linear perturbations of the scalar fields do not exhibit tachyonic instabilities whose time scales are parametrically shorter than the cosmological time scale. Under the condition (\ref{eqn:condition-for-entropy-bound}), one can safely apply the covariant entropy bound, leading to the upper bound on the entropy of the system $S \leq \pi/H^2$. By combining this with (\ref{eqn:entropy-ansatz}) and setting $R\sim 1/H$, one obtains $N \lesssim H^{-(2-\delta_2)/\delta_1}$ under the condition (\ref{eqn:condition-for-entropy-bound}). As argued in ref.~\cite{Ooguri:2018wrx} it is expected that this upper bound on $N$ should be an increasing function of the horizon radius and thus is a decreasing function of $H$, namely $(2-\delta_2)/\delta_1>0$. Motivated by the fact that Bekenstein bound tends to saturate for large $N$, ref.~\cite{Ooguri:2018wrx} further assumes that the covariant entropy bound in this form also saturates for large $N$. As a result, one obtains $S\sim S_{\rm tower}(N,R)$ and 
\begin{equation}
 N \sim \left(\frac{1}{H}\right)^{\frac{2-\delta_2}{\delta_1}}\,, \quad \delta_1 > 0\,, \quad 0 < \delta_2 < 2\,, \label{eqn:entropy-bound-saturated}
\end{equation}
for $H\ll 1$ under the condition (\ref{eqn:condition-for-entropy-bound}).

While the right hand side of (\ref{eqn:N-ansatz}) is a function of $\phi$, that of (\ref{eqn:entropy-bound-saturated}) is a function of time. When $\partial_{\mu}\phi$ is timelike during the cosmological evolution of the scalar fields, $\phi$ can be considered as a time variable. From (\ref{eqn:N-ansatz}) and (\ref{eqn:entropy-bound-saturated}) one then obtains $\ln n(\phi) \sim -b\phi - \frac{2-\delta_2}{2\delta_1}\ln H^2$. Plugging this to the inequality in (\ref{eqn:N-ansatz}) results in 
\begin{equation}
 \left|\frac{1}{H^2}\frac{d(H^2)}{d\phi}\right| \gtrsim c_0\,, \quad c_0 \equiv \frac{2b\delta_1}{2-\delta_2}\,. 
  \label{eqn:outsome-of-entropy-bound}
\end{equation}
We have thus shown that (\ref{eqn:condition-for-entropy-bound}) implies (\ref{eqn:outsome-of-entropy-bound}). Conversely, if (\ref{eqn:outsome-of-entropy-bound}) is violated then at least one of the inequalities in (\ref{eqn:condition-for-entropy-bound}) must be violated. It is therefore concluded that
\begin{equation}
 \left|\frac{1}{H^2}\frac{d(H^2)}{d\phi}\right| \gtrsim c_0\,, \quad \mbox{or}\quad
 \left|\frac{\dot{H}}{H^2}\right| \gtrsim c_1\,, \quad \mbox{or}\quad
  \frac{\min m_{\rm scalar}^2}{H^2} \lesssim -c_2\,,  \label{eqn:conjecture}
 \end{equation}
 where $H$ and $\dot{H}$ are the Hubble expansion rate and its derivative with respect to the proper time in the Einstein frame,  $\phi$ is the geodesic distance from a point in the field space deep inside the regime of validity of the effective field theory, $\min m_{\rm scalar}^2$ is the lowest among squared masses of perturbation modes of the scalar fields, and $c_{0,1,2}$ are positive numbers of $\mathcal{O}(1)$. The condition (\ref{eqn:conjecture}) is the refined de Sitter swampland conjecture rewritten in a way that is convenient for extensions. For slow-roll models with canonical kinetic terms, the conjecture (\ref{eqn:conjecture}) indeed reduces to (\ref{eqn:conjecture-slowroll}), where $c\equiv \min(c_0, \sqrt{2c_1})$ and $c'\equiv c_2/3$ are still of $\mathcal{O}(1)$.

We now make a couple of comments on the computation of the quantity $\min m_{\rm scalar}^2$ in the conjecture (\ref{eqn:conjecture}). First, mixing between the perturbations of the scalar fields and the metric perturbations lead to corrections to $\min m_{\rm scalar}^2$ of $\mathcal{O}(H^2)$. However, assuming that $c_2$ is not too small, we can ignore such corrections. Therefore, one can compute $\min m_{\rm scalar}^2$ in the decoupling limit, i.e. in the $M_{\rm Pl}^2\to\infty$ limit. Second, as we shall see in the next section, when the field space is curved, $\min m_{\rm scalar}^2$ in the decoupling limit can be significantly different from the lowest eigenvalue of the Hessian of the potential.

The refined de Sitter swampland conjecture appears to be a serious challenge to the inflationary scenario of the early universe and dark energy models of the late-time universe based on scalar fields, provided that $c_{0,1,2}$ in (\ref{eqn:conjecture}) are really of $\mathcal{O}(1)$, which is supported by some concrete compactifications in \cite{Obied:2018sgi} but never proven theoretically. On the other hand, in our universe, some constants that are expected to be of $\mathcal{O}(1)$ from theoretical viewpoints may turn out to be rather small or large. A well-known example is the cosmological constant: a theoretically natural value of the cosmological constant in the unit of $M_{\rm Pl}$ is of $\mathcal{O}(1)$ while observations tell that it should actually be as small as $10^{-120}$. Therefore, even if the conjecture is right, the ``$\mathcal{O}(1)$'' numbers $c_{0,1,2}$ may turn out to be small in our universe so that the low energy limit of a consistent theory of quantum gravity may accommodate models of inflation in the early universe and/or dark energy modes in the late-time universe based on scalar fields. If this is the case then understanding the nature and the origin of the smallness of those numbers would advance our knowledge in theoretical physics. Towards this goal, it is important to see how far the conjecture can be extended in a consistent way. Otherwise, the conjecture would be applicable to only a small subset of models of inflation and dark energy and would not tell anything about models outside the regime of applicability. In the following we shall initiate such an extension and we hope that our approach will be useful for further extensions.

 \section{Two-field model with  hyperbolic field space}
 \label{secHS}

 The distance conjecture indicates that the geometry of the moduli space should be negatively curved, at least in the vicinity of the infinity~\cite{Ooguri:2006in}. It is therefore worthwhile considering a negatively curved field space in the context of the de Sitter swampland conjecture.

 As the simplest case, we consider a two-dimensional hyperbolic field space,  
 \begin{equation}
  \mathcal{G}_{IJ}(\phi^K)d\phi^Id\phi^J = d\chi^2 + e^{2\beta\chi}d\varphi^2\,,
 \end{equation}
 where $\beta$ is a positive constant. For concreteness and for the reason that will be clarified soon, we suppose that the potential is a linear combination of $\cosh\left[2\beta(\chi-\chi_0)\right]$ and $1$ with $\varphi$-dependent coefficients, i.e. $V(\chi,\varphi) = A(\varphi)\cosh\left[2\beta(\chi-\chi_0)\right] + B(\varphi)$, where $\chi_0$ is a constant. By shifting the origin of $\chi$ and rescaling $\varphi$, one can set $\chi_0=0$ without loss of generality. The action of the scalar fields $\{\phi^I\}=\{\chi,\varphi\}$ is then given by 
 \begin{equation}
  I = \int d^4x\sqrt{-g}\left\{ -\frac{1}{2}g^{\mu\nu}\partial_{\mu}\chi\partial_{\nu}\chi - \frac{1}{2}e^{2\beta\chi}g^{\mu\nu}\partial_{\mu}\varphi\partial_{\nu}\varphi - T(\varphi)\left[\cosh(2\beta\chi)-1\right] - U(\varphi) \right\}\,,
   \label{eqn:action-twofields}
 \end{equation}
where $T(\varphi)\equiv A(\varphi)$ and $U(\varphi)\equiv A(\varphi)+B(\varphi)$. This model is coupled to the Einstein gravity in four dimensions.

\subsection{Attractor behavior of $\chi$}
\label{subsec:attractor}

The equation of motion for $\chi$ leads
\begin{equation}
 \Box\chi + 2\beta e^{2\beta\chi}X - 2\beta T(\varphi)\sinh(2\beta\chi) = 0\,, \label{eqn:eom-chi}
\end{equation}
where $X\equiv -g^{\mu\nu}\partial_{\mu}\varphi\partial_{\nu}\varphi/2$. As we shall see below, if $\beta^2$ is large then $\chi$ has a heavy mass around the value determined by the equation of motion (\ref{eqn:eom-chi}) with $\Box\chi$ dropped, i.e. 
\begin{equation}
 2\beta e^{2\beta\chi}X - 2\beta T(\varphi)\sinh(2\beta\chi) \simeq 0\,,\label{eqn:eom-chi2}
\end{equation}
which is easily solved with respect to $\chi$ as
\begin{equation}
 \chi \simeq \frac{1}{2\beta}\ln\gamma\,, \quad \gamma \equiv \left(1-\frac{2X}{T(\varphi)}\right)^{-1/2}\,.
  \label{eqn:chi-value}
\end{equation}
The second derivative of the potential with respect to $\chi$ around this time-dependent value is
\begin{equation}
 \partial_{\chi}^2 V |_{2\beta\chi=\ln\gamma} = \frac{4T}{\gamma}\beta^2 \,, 
\end{equation}
and it can be made arbitrarily large by setting $\beta^2$ large enough. Therefore, for large enough $\beta^2$ it is expected that the trajectory (\ref{eqn:chi-value}) is an attractor of the system, that the two-dimensional field space is reduced to an effective one-dimensional field space spanned by $\varphi$ and that the system is described by the effective action, 
\begin{equation}
 I_{\rm eff} = \int d^4x\sqrt{-g}\left\{T(\varphi)\left[ -\sqrt{1-\frac{2X}{T(\varphi)}} + 1\right] - U(\varphi)\right\}\,, \quad
  X = -\frac{1}{2}g^{\mu\nu}\partial_{\mu}\varphi\partial_{\nu}\varphi\,,
  \label{eqn:Ieff}
\end{equation}
which is obtained by simply substituting (\ref{eqn:chi-value}) to (\ref{eqn:action-twofields}). This is nothing but a DBI action. Actually, we have chosen the form of the potential $V(\chi,\varphi)$ so that the effective single-field theory after integrating out the heavy field is described by the DBI action. This is a special case of the gelaton scenario originally proposed in~\cite{Tolley:2009fg} and further extended in \cite{Elder:2014fea}.

\begin{figure}[ht]
\centering
\includegraphics[scale=0.5]{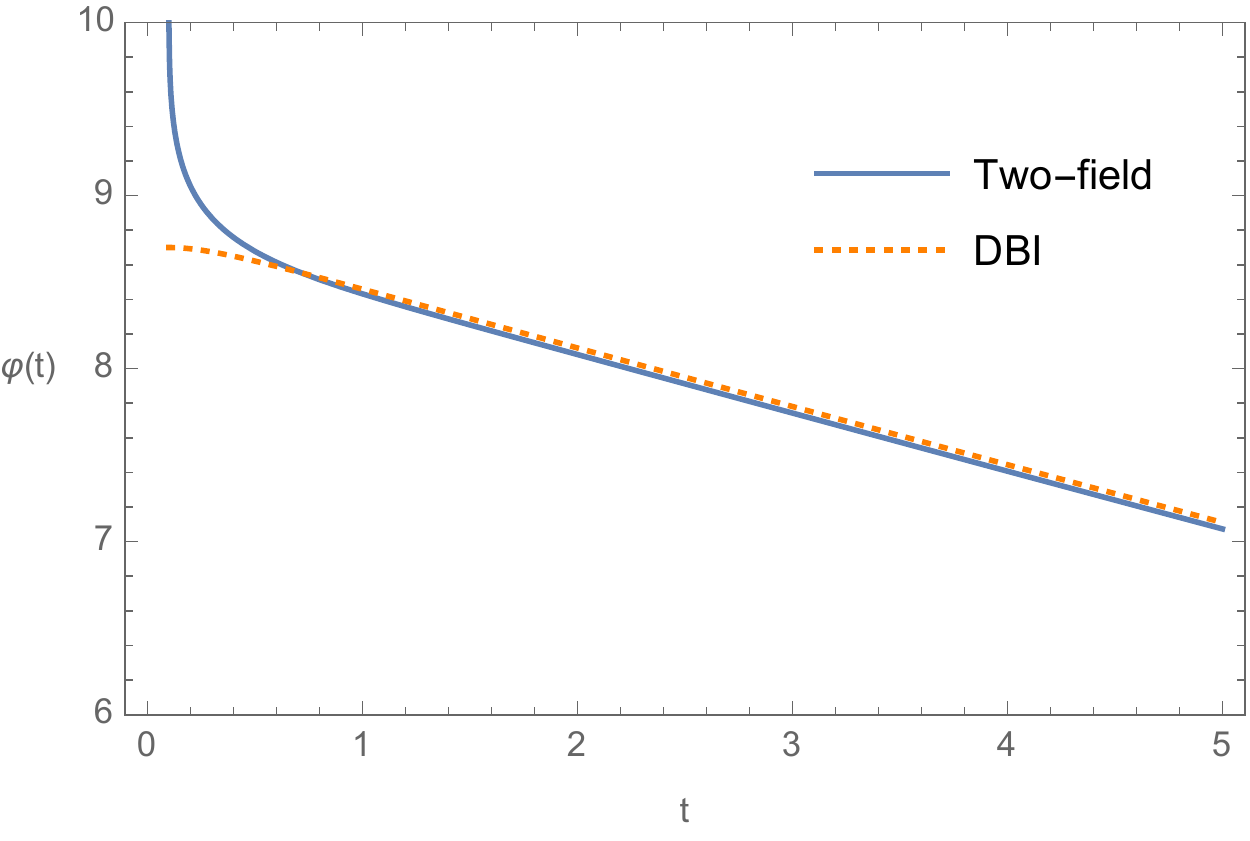}\qquad\qquad
\includegraphics[scale=0.5]{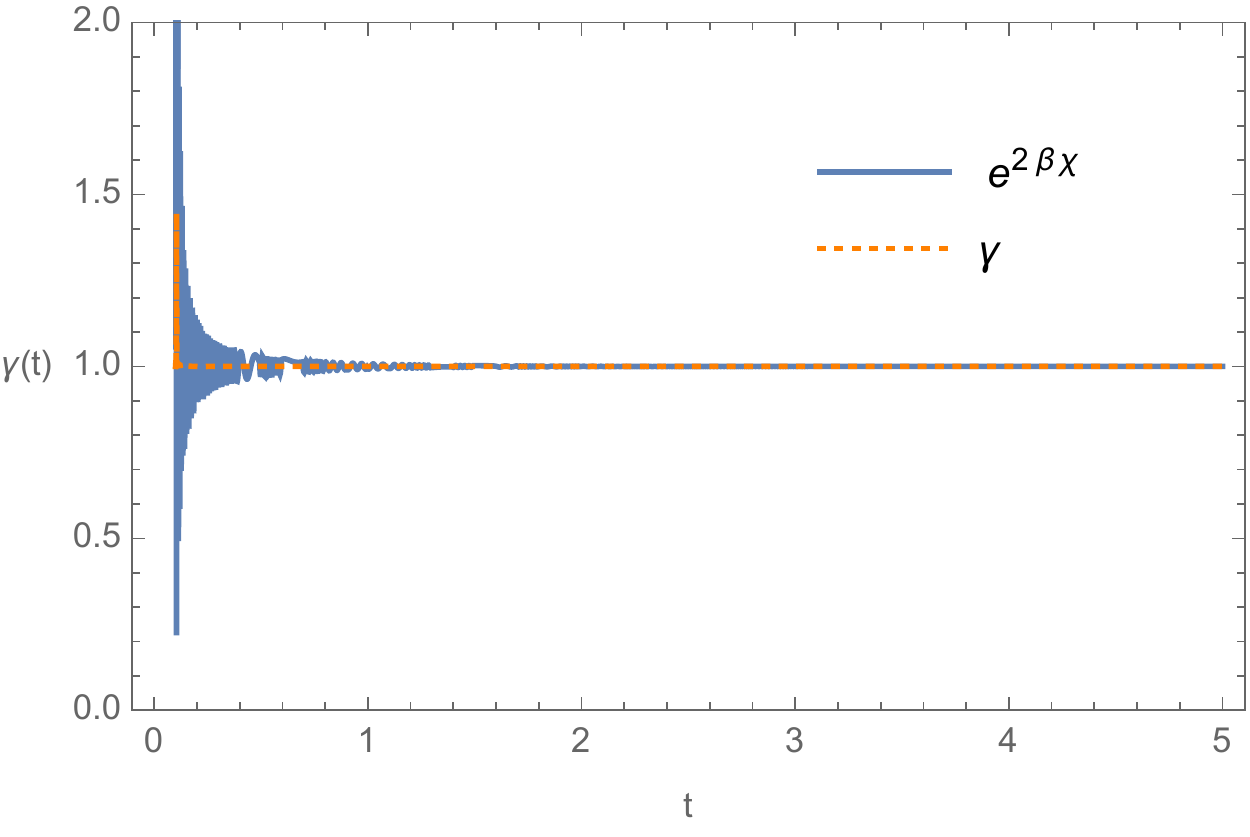}
\caption{Non-relativistic attractor $(\gamma=1)$ with the parameter choice $U(\varphi)=1+ 0.1\varphi^2$, $T(\varphi)\equiv \varphi^4/\lambda$, $\beta=20$ and $\lambda=0.5$.\label{fig1}
}
\end{figure}

\begin{figure}[ht]
\centering
\includegraphics[scale=0.5]{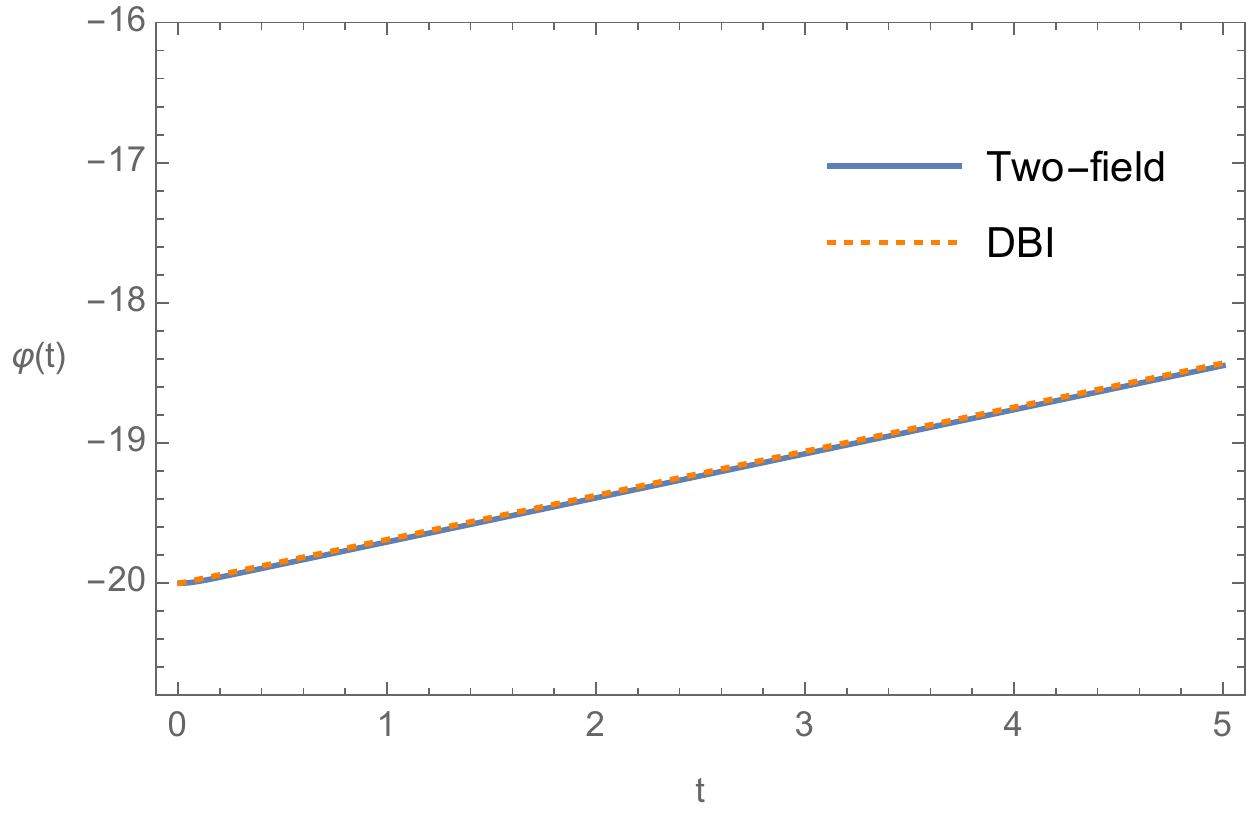}\qquad\qquad~
\includegraphics[scale=0.5]{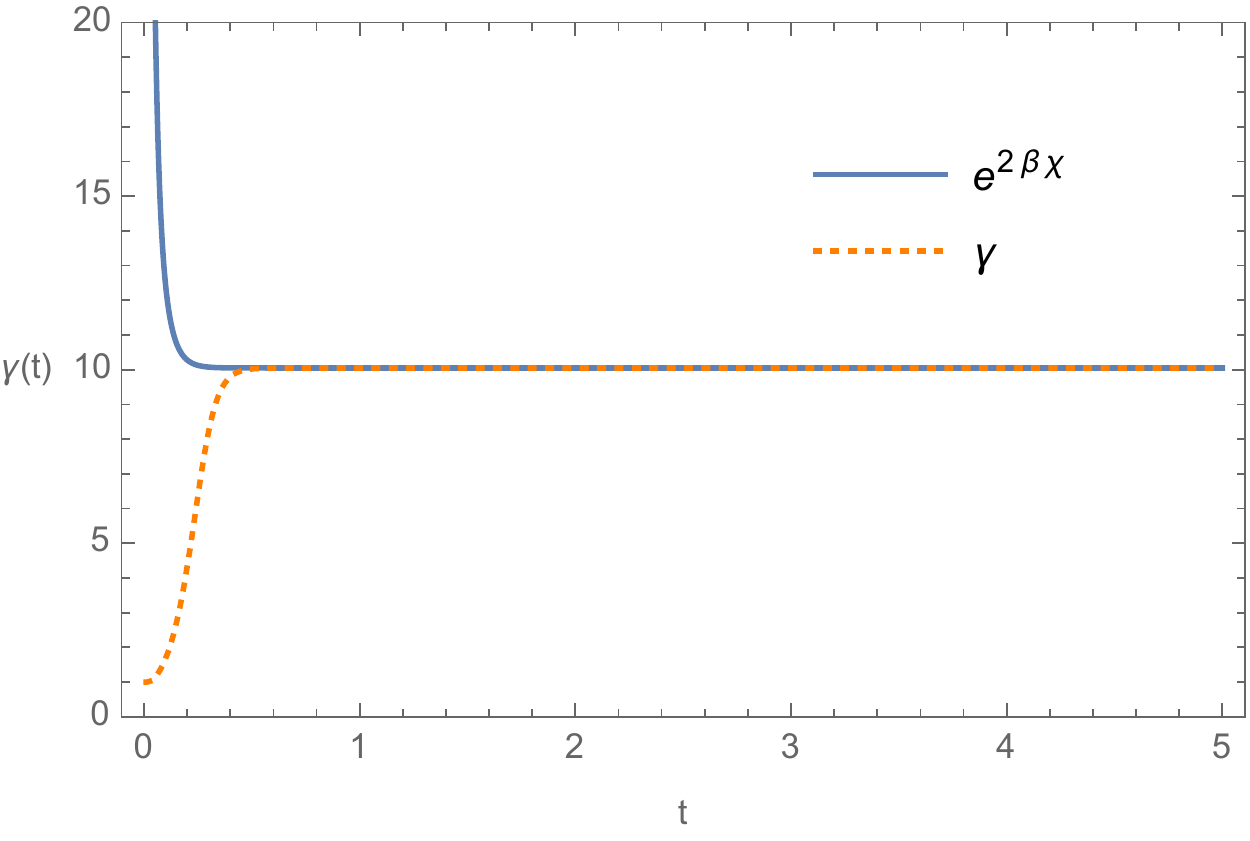}
\caption{Relativistic attractor $(\gamma=10)$ with the
parameter choice $U(\varphi)=7.5\varphi^2$, $T(\varphi)\equiv1/\lambda$,  $\beta=20$ and $\lambda=10$. \label{fig2}
}
\end{figure}

For some choices of the functions $T(\varphi)$ and $U(\varphi)$, we have numerically confirmed that the trajectory (\ref{eqn:chi-value}) is an attractor of the system, that the deviation of the system from (\ref{eqn:chi-value}) quickly decays, and that the system is well described by the single-field action (\ref{eqn:Ieff}), provided that $\beta^2$ is large enough. See Fig. \ref{fig1} for an example where the attractor is non-relativistic ($\gamma=1$), while see Fig. \ref{fig2} for an example where the attractor is relativistic ($\gamma=10$) \footnote{Actually, once we fix the combination of $T(\varphi)$ and $U(\varphi)$ in the DBI model, we can classify the late-time attractor, see \cite{Copeland:2010jt}.}. In the left plots, ``two-field'' denotes that we evaluate $\varphi (t)$ based on the full equations of motion from the two-field action (\ref{eqn:action-twofields})
\begin{align}
&\ddot{\chi} +3H\dot{\chi} - {\beta}e^{2\beta\chi} \dot{\varphi}^2 + 2 \beta T(\varphi) \sinh{(2\beta\chi)}=0\,,\nonumber \\
&\ddot{\varphi} +3H\dot{\varphi}+  2\beta\dot{\chi}\dot{\varphi} +\frac{T'(\varphi)}{e^{2\beta\chi}} \big[ \cosh(2\beta\chi)-1\big]+\frac{U'(\varphi)}{e^{2\beta\chi}}=0\,,\nonumber \\
&3H^2 =\frac{1}{2}(\dot{\chi}^2+ e^{2\beta\chi} \dot{\phi}^2)+ {T(\varphi)}\big[\cosh (2\beta\chi) -1 \big]+ {U(\varphi) }\,,
\label{eqn:two}
\end{align}
while ``DBI'' denotes that we evaluate $\varphi (t)$ based on the equation of motion for the single-field DBI model
\begin{align}
&{\gamma}^2\ddot{\varphi} + {3 H} \dot{\varphi}  - \frac{T'(\varphi)}{2} \frac{(\gamma -1)^2(\gamma+2) }{\gamma}  +\frac{U'(\varphi)}{\gamma}=0\,, \nonumber \\
&3H^2 =\frac{\gamma^2}{(\gamma+1)}\dot{\varphi}^2+ {U(\varphi) } , \qquad 
\gamma = \left(1-\frac{\dot{\varphi}^2}{T(\varphi)}\right)^{-1/2} \, . 
\end{align}
In the right plots, we confirm that the attractors satisfying the relation (\ref{eqn:chi-value}), i.e. $\gamma=e^{2\beta\chi}$, are realized quickly. In these plots, both of ``$e^{2\beta\chi}$'' and``$\gamma$'' are evaluated based on the full equations of motion \eqref{eqn:two} from the two-field action. 

\subsection{Geodesic distance in the field space}

The first inequality in the conjecture (\ref{eqn:conjecture}) involves the geodesic distance $\phi$ from a point in the field space deep inside the regime of validity of the effective field theory. In the model (\ref{eqn:action-twofields}), it is given by integrating
\begin{equation} 
 d\phi = \sqrt{\mathcal{G}_{IJ}(\phi^K)d\phi^I d\phi^J} = \sqrt{d\chi^2 + e^{2\beta\chi}d\varphi^2}\,. 
\end{equation}
For large enough $\beta^2$, by using the attractor behavior (\ref{eqn:chi-value}), this is reduced to
\begin{equation}
 d\phi \simeq  \left(\frac{\dot{\gamma}^2}{4\beta^2\gamma^2\dot{\varphi}^2} + \gamma \right)^{1/2}
  d\varphi \simeq \sqrt{\gamma}d\varphi\,, \label{eqn:dphi-largebeta}
\end{equation}
where we have used the fact that the evolution of $\varphi$ is well described by the single-field model and thus $\dot{\gamma}^2/(\gamma^2\dot{\varphi}^2)$ remains finite in the $\beta^2\to\infty$ limit. Thus, the first inequality in the conjecture (\ref{eqn:conjecture}) can be rewritten as
\begin{equation}
  \frac{1}{\sqrt{\gamma}}\left|\frac{1}{H^2}\frac{d(H^2)}{d\varphi}\right| \gtrsim c_0\,. 
\end{equation}
It is interesting to see that other quantities associated with $\chi$ in the original two-field model do not appear in this inequality.

\subsection{Squared masses of scalar perturbation modes} 

The last inequality in the conjecture (\ref{eqn:conjecture}) involves the lowest squared mass $\min m_{\rm scalar}^2$ of perturbation modes of the scalar fields. As already stated, one is allowed to take the decoupling limit, i.e. the $M_{\rm Pl}^2\to\infty$ limit, to simplify the computation. Even in this limit, the field space is highly curved for large $\beta^2$ and the curved field space makes the computation non-trivial. As we shall see explicitly, the leading contribution to $\min m_{\rm scalar}^2$ in the decoupling limit does not agree with the lowest eigenvalue of the Hessian of the potential~\footnote{In the language of the covariant formalism~\cite{Sasaki:1995aw}, one needs to take into account not only the covariant version of the Hessian of the potential and the field-space Riemann tensor contracted twice with the time derivative of the background scalar fields but also the connection terms that are involved in the covariant time derivatives of perturbations.}. The difference comes from e.g. the friction terms of order $\mathcal{O}(\beta)$ that mix $\delta\chi$ and $\delta\varphi$.

We consider general linear perturbations around a homogeneous and isotropic background with the scale factor $a(t)$ and then decompose them into scalar, vector and tensor parts as usual. The action quadratic in perturbations for scalar, vector and tensor parts are decoupled from each other and thus can be analyzed separately. Therefore, for the purpose of writing down the last inequality of (\ref{eqn:conjecture}), we just need to consider the scalar sector. By adopting the spatially flat gauge, the scalar fields and the metric in the Einstein-frame are then written as 
\begin{eqnarray}
 \varphi & = & \varphi^{(0)}(t) + \delta\varphi(t,\vec{x})\,,\nonumber\\
 \chi & = & \chi^{(0)}(t) + \delta\chi(t,\vec{x})\,,\nonumber\\
 g_{\mu\nu}dx^{\mu}dx^{\nu} & = & - \left[1+2\Phi(t,\vec{x})\right] dt^2 + 2 N(t)a(t)\partial_iB(t,\vec{x}) dtdx^i + a(t)^2 \delta_{ij} dx^idx^j\,. 
\end{eqnarray}
Expanding the total action up to second order in perturbations and performing some integrations by part, it is found that the action does not contain time derivatives of $\Phi$ and $B$. We then integrate out $\Phi$ and $B$ from the quadratic action by using their equations of motion. After substituting the attractor solution 
\begin{equation}
 \chi^{(0)} \simeq \frac{1}{2\beta}\ln\gamma^{(0)}\,, \quad \gamma^{(0)} \equiv \left(1-\frac{(\dot{\varphi}^{(0)})^2}{T(\varphi^{(0)})}\right)^{-1/2} \label{eqn:background-attractor}
\end{equation}
at the background level~\footnote{For the time derivative of $\chi^{(0)}$, we substitute $\dot{\chi}^{(0)} = \dot{\gamma}^{(0)}/(2\beta\gamma^{(0)})$} and performing some integrations by part, we obtain the quadratic part of the action \eqref{eqn:action-twofields} as
\begin{equation}
 I^{(2)} = \frac{1}{2}\int dt a^3\left[ \dot{Y}^{\rm T}\mathcal{K}\dot{Y} + \dot{Y}^{\rm T}\mathcal{M}Y +Y^{\rm T}\mathcal{M}^\text{T}\dot{Y} - Y^{\rm T}\left(-\mathcal{K} \frac{\vec{\nabla}^2}{a^2} + \mathcal{V}\right)Y\right]\,, \quad Y = \left(\begin{array}{c}
					    \delta\varphi \\
																								    \delta\chi
																								   \end{array}\right)\,,
\end{equation}
where
\begin{equation}
 \mathcal{K} = \left(\begin{array}{cc}
		 \gamma & 0\\
		      0 & 1
		       \end{array}\right)\,, \quad
 \mathcal{M} = \left(\begin{array}{cc}
		0 & \beta\gamma\dot{\varphi} \\
		    -\beta\gamma\dot{\varphi}   & 0
		       \end{array}\right)\,, \quad
 \mathcal{V} = \left(\begin{array}{cc}
		\mathcal{V}_{11} & \mathcal{V}_{12}\\
		      \mathcal{V}_{12} & \mathcal{V}_{22}
		       \end{array}\right) + \mathcal{O}(M_{\rm Pl}^{-2}) \,,
 \label{eqn:def-MKV}
\end{equation}
and 
\begin{eqnarray}
\mathcal{V}_{11}   =   \frac{(\gamma -1 )^2}{2\gamma}  T'' + U'' \,,\quad
\mathcal{V}_{12}   =   \frac{(\gamma +3)(\gamma -1)}{2\gamma} T' \beta- U' \beta \,, \quad
\mathcal{V}_{22}   =   \frac{4T}{\gamma}\beta^2\,.
\end{eqnarray}
Hereafter we suppress the superscript $(0)$ for the background quantities. Assuming that the time scale of the evolution of each component of the three matrices $\mathcal{K}$, $\mathcal{M}$ and $\mathcal{V}$ is cosmological, i.e. of $\mathcal{O}(H^{-1})$, the leading contributions~\footnote{We can ignore the corrections due to the cosmological time-dependence of the components of the three matrices, assuming that $c_2$ is not too small.} to the squared masses $m_{\rm scalar}^2$ for scalar perturbations are obtained by solving the following second-order algebraic equation for $m^2$,
\begin{equation}
 \det \left[ m^2 \mathcal{K} - 2im\mathcal{M} - \mathcal{V} \right] = 0\,.
\end{equation}
There are two independent solutions $m^2=m_{\pm}^2$, where
\begin{eqnarray}
 m_+^2 & = & 4T(\varphi) \gamma \beta^2 + \mathcal{O}(\beta^0)\,, \nonumber\\
 m_-^2 & = & \Omega  + \mathcal{O}(\beta^{-2}) + \mathcal{O}(M_{\rm Pl}^{-2})\,, \label{eqn:mpm2}
\end{eqnarray}
and 
\begin{equation}
\Omega = \frac{1}{\gamma^3}U'' + \frac{(\gamma-1)^2}{2\gamma^4} T'' - 
\frac{\left[( \gamma +3)(\gamma-1) T' - 2\gamma  U' \right]^2\,}{16\gamma^4 T}\,. \label{eqn:Omega}
\end{equation}
An alternative derivation can be found in the Appendix \ref{AppA}. Since $m_+^2\to +\infty$ in the $\beta^2\to\infty$ limit, the last inequality in the conjecture (\ref{eqn:conjecture}) is reduced to
\begin{equation}
   \frac{\Omega}{H^2} \lesssim -c_2\,. \label{eqn:thirdinequality-largebeta}
\end{equation}
The precise expression of $\Omega$ depends on the choice of perturbation variables and the gauge. However, the difference is of $\mathcal{O}(H^2)$ and thus does not matter, provided that $c_2$ is not too small.

 \section{Swampland conjecture for DBI scalar\label{secDBI}}

 In the previous section we showed that $\chi$ in the two-field model (\ref{eqn:action-twofields}) becomes infinitely heavy in the $\beta^2\to\infty$ limit and the evolution of the full system is well described by the attractor solution (\ref{eqn:chi-value}) for $\chi$ and the single-field DBI action
\begin{equation}
 I_{\rm DBI} = \int d^4x\sqrt{-g}\left\{T(\varphi)\left[ -\sqrt{1-\frac{2X}{T(\varphi)}} + 1\right] - U(\varphi)\right\}
  \label{eqn:DBIaction}
\end{equation}
for $\varphi$. We have also shown that for the two-field model with large $\beta^2$, the swampland conjecture (\ref{eqn:conjecture}) is written as 
\begin{equation}
 \frac{1}{\sqrt{\gamma}}\left|\frac{1}{H^2}\frac{d(H^2)}{d\varphi}\right| \gtrsim c_0\,, \quad \mbox{or}\quad
 \left|\frac{\dot{H}}{H^2}\right| \gtrsim c_1\,, \quad \mbox{or}\quad
   \frac{\Omega}{H^2} \lesssim -c_2\,, \label{eqn:conjecture-DBI}
 \end{equation}
 where $\gamma \equiv 1/\sqrt{1-2X/T}$, $X \equiv -g^{\mu\nu}\partial_{\mu}\varphi\partial_{\nu}\varphi/2$ and
\begin{equation}
 \Omega = \frac{1}{\gamma^3}U'' + \frac{(\gamma-1)^2}{2\gamma^4} T'' 
- \frac{\left[( \gamma +3)(\gamma-1) T' - 2\gamma  U' \right]^2\,}{16\gamma^4 T} \,. 
  \label{eqn:def-Omega}
\end{equation}
It is obvious that all quantities appearing in (\ref{eqn:conjecture-DBI})-(\ref{eqn:def-Omega}) are well-defined within the single-field DBI model (\ref{eqn:DBIaction}) coupled to four-dimensional general relativity and do not rely on any quantities that are defined only in the original two-field model \eqref{eqn:action-twofields} such as $\beta$.

We thus speculate that the condition (\ref{eqn:conjecture-DBI})-(\ref{eqn:def-Omega}) derived in the infinite curvature limit should be considered as the swampland conjecture for a DBI scalar field by its own. This is consistent with the very idea of the swampland conjecture that states necessary conditions for effective field theories to be UV-completable: such a conjecture can be useful only if those conditions are written in terms of quantities well-defined within effective field theories. Suppose that there are two single-field DBI models that are completely identical (i.e. having the same $T(\varphi)$ and $U(\varphi)$) at the level of low energy effective field theories, that one of them is merely the effective single-field description of the two-field model (\ref{eqn:action-twofields}) and that the other describes e.g. the motion of a D-brane in extra dimensions. From the viewpoint of low-energy effective field theories, there is no way to distinguish these two models and there are no quantities within the effective field theories that differ between the two models. Therefore, if there is a swampland conjecture for single-field DBI models at all then the conditions that the conjecture imposes on the first model should be exactly the same as the conditions on the second model. Hence the consistency argument requires that (\ref{eqn:conjecture-DBI})-(\ref{eqn:def-Omega}) should be the universal condition that the swampland conjecture imposes on all single-field DBI models.

As yet another consistency check, let us now consider the limit where the single-field DBI model is reduced to a scalar field model with a canonical kinetic term and the potential $U$, namely the $X/T\to 0$ limit. In this limit the quantities $X$ and $U$ need to be well defined and thus should be kept finite. Hence we express $T$ and $\gamma$ as $T=2X/(1-1/\gamma^2)$ and $\gamma = 1 + \gamma_1$, consider $\gamma_1$ as a small quantity and calculate the leading contribution to $\Omega$ in the $\gamma_1\to 0$ limit. The result is
\begin{equation}
 \Omega = U'' + \mathcal{O}(\gamma_1)\,,
\end{equation}
where we have kept $(\ln T)'$ and $(\ln T)''$ kept finite in the limit. Thus (\ref{eqn:conjecture-DBI})-(\ref{eqn:def-Omega}) correctly recovers
\begin{equation}
 \left|\frac{1}{H^2}\frac{d(H^2)}{d\varphi}\right| \gtrsim c_0\,, \quad \mbox{or}\quad
 \left|\frac{\dot{H}}{H^2}\right| \gtrsim c_1\,, \quad \mbox{or}\quad
   \frac{U''}{H^2} \lesssim -c_2\,, \quad (\mbox{canonical limit}). 
 \end{equation} 
In the slow roll case this of course reduces to (\ref{eqn:conjecture-slowroll}) with $V$ replaced by $U$ and $\phi$ replaced by $\varphi$. 
 
\section{Swampland conjecture for $P(X,\varphi)$ theories\label{secPX}}

More generally, following the logic that we have proposed, it is straightforward to extend the de Sitter swampland conjecture to a general k-essence type scalar field with the Lagrangian $P(X,\varphi)$, where $X \equiv - g^{\mu\nu}\partial_{\mu}\varphi \partial_{\nu}\varphi/2$. First, the Lagrangian is equivalent to $L = P(\hat\chi,\varphi) + \lambda ( X -\hat\chi )$, where $\lambda$ is a Lagrange multiplier. Second, one can eliminate $\lambda$ by using the equation of motion for $\hat\chi$ as $L = P(\hat\chi,\varphi) + P_{,\hat\chi}(\hat\chi,\varphi)(X - \hat\chi)$. Third, one can then deform the Lagrangian by giving a tiny kinetic term to $\hat\chi$ as $\tilde{L} = L - \frac{Z^2}{2} g^{\mu\nu}\partial_{\mu}\hat\chi \partial_{\nu}\hat\chi $, where $Z$ is a small constant~\cite{Elder:2014fea}. 
The total action is
 \begin{equation}
  I = \int d^4x\sqrt{-g}\left\{ -\frac{Z^2}{2}g^{\mu\nu}\partial_{\mu}\hat\chi\partial_{\nu}\hat\chi + P_{,\hat\chi}(\hat\chi,\varphi)(X - \hat\chi) +P(\hat\chi,\varphi) \right\}.
   \label{eqn:action-PX}
 \end{equation}
This two-field system has the field space metric of the form $P_{,\hat\chi}(\hat\chi,\varphi)d\varphi^2 + Z^2d\hat\chi^2$ and thus the geodesic distance in the field space is $d\phi = \sqrt{P_{,\hat\chi}(\hat\chi,\varphi)+Z^2(d\hat\chi/d\varphi)^2}d\varphi$. By taking the $Z\to 0$ limit, it is concluded that we should use $d\phi = \sqrt{P_{,X}(X,\varphi)}d\varphi$ in the original single-field system with the Lagrangian $P(X,\varphi)$.

It is straightforward to analyze scalar perturbations around a flat FLRW background in the decoupling limit and to see that in the zero momentum sector there are two fast modes $\sim e^{\pm m_+ t}$ with $m_{+}^2 = \mathcal{O}(Z^{-2})$ and two slow modes $\sim e^{\pm m_- t}$ with $m_{-}^2 = \mathcal{O}(Z^0)$. Obviously, the stability of the two-field system requires $m_{+}^2>0$ and hence we have $\min m_{\rm scalar}^2 = m_{-}^2$. We thus end up with the following de Sitter swampland conjecture for a scalar field described by the Lagrangian $P(X,\varphi)$, 
\begin{equation}
 \frac{1}{\sqrt{P_{,X}(X,\varphi)}}\left|\frac{1}{H^2}\frac{d(H^2)}{d\varphi}\right| \gtrsim c_0\,,
  \quad \mbox{or}\quad
   \left|\frac{\dot{H}}{H^2}\right| \gtrsim c_1\,, \quad \mbox{or}\quad
  \frac{m_{-}^2}{H^2} \lesssim -c_2\,.  \label{eqn:conjecture-kessence}
\end{equation}
The precise expression of $m_-^2$ depends on the choice of perturbation variables and the gauge. However, the difference is of $\mathcal{O}(H^2)$ and thus does not matter, provided that $c_2$ is not too small.

As a special case, we can choose $P(X,\varphi)$ as the Lagrangian in the DBI action \eqref{eqn:DBIaction}, which means $P(X,\varphi)=T(\varphi)\left[ -\sqrt{1-\frac{2X}{T(\varphi)}} + 1\right] - U(\varphi)$ and $P_{,X}(X,\varphi)=1/\sqrt{1-\frac{2X}{T(\varphi)}}\equiv\gamma$.
Then \eqref{eqn:conjecture-kessence} will recover the swampland conjecture for a DBI scalar field in \eqref{eqn:conjecture-DBI}.
Actually, in the $Z\to 0$ limit, the attractor solution from \eqref{eqn:action-PX} is $\hat\chi\simeq X=\frac{T}{2}\left(1- \gamma^{-2}\right)$ and thus the geodesic distance in the field space is $d\phi\simeq \sqrt{\gamma}d\varphi$ as in (\ref{eqn:dphi-largebeta}). The analysis of the linear perturbation around a homogeneous and isotropic background shows that in the small $Z$ limit the scalar sector of this two-field model also consists of fast modes with the squared mass of order $\mathcal{O}(Z^{-2})$ and slow modes with the squared mass of order $\mathcal{O}(Z^0)$. The last inequality in the conjecture (\ref{eqn:conjecture}) thus remains well-defined in the limit and gives essentially the same condition as (\ref{eqn:thirdinequality-largebeta}) up to terms of order $\mathcal{O}(H^2)$ in $\Omega$ that are unimportant for not too small $c_2$.

\section{Summary and discussion\label{secCon}}

Recently, the swampland conjectures,  that may constrain low energy effective field theories from the viewpoint of whether they admit consistent UV completion with gravity,  have attracted much attention. Especially, by applying the de Sitter swampland conjecture to inflation and quintessence, many new constraints that cannot be obtained just by observations have been found.  Regardless of this, most discussions of the application of the de Sitter swampland conjecture to scalar field(s) were limited to models with linear kinetic terms of the form (\ref{eqn:Lkin}). Although there are some preceding works, where its applications to inflation models with nonlinear kinetic terms were discussed~\cite{Kehagias:2018uem,Seo:2018abc,Bhattacharya:2018xlw,Herdeiro:2018hfp,Heisenberg:2019qxz}, it is fair to say that the discussion has not been settled down. Therefore, in this paper, we tried to establish a plausible way to apply the de Sitter swampland conjecture to inflation models with nonlinear kinetic terms, with a special emphasis on DBI model. 

Our method to obtain the de Sitter swampland conjecture for DBI model includes the following steps. Firstly, we summarize the recently proposed refined de Sitter swampland conjecture for scalar fields with linear kinetic terms that is derived from the combination of the distance conjecture and the Bousso's entropy bound. Then we consider a two-field model with a hyperbolic field space, where in the infinitely curved limit one field $\chi$ is infinitely heavy and trapped at the minimum. Because of this attractor behavior, by integrating out the degree of freedom for $\chi$, we can obtain an effective single-field model that includes only the other field $\varphi$. We show that the single-field DBI model can be obtained by appropriately choosing the form of the two-field potential within this scheme. Finally, we apply the de Sitter swampland conjecture to the two-field model in the infinitely curved limit which is equivalent to the single-field DBI model and obtain the conditions given by  \eqref{eqn:conjecture-DBI}-\eqref{eqn:def-Omega}.

We show that the quantities related with the de Sitter swampland conjecture for the two-field model, like the geodesic distance in the field space and squared mass of scalar perturbation modes can be also well-defined in terms of the quantities of the single-field DBI model in the infinitely curved limit. This fact suggests that we can regard the de Sitter swampland conjecture for this set-up as the one for the single-field DBI model by its own. The de Sitter swampland conjecture for the DBI model has been also discussed in \cite{Seo:2018abc}, where the proper distance in the AdS bulk is identified with the scalar field related with the de Sitter swampland conjecture. However, with this choice, it is found that in the non-relativistic regime, the conjecture becomes not consistent with the Bousso's entropy bound. Since in our approach, where the geodesic distance in the field space is relevant to the de Sitter swampland conjecture, the conjecture keeps to be consistent with the Bousso's entropy bound even in the non-relativistic regime, we speculate that our approach is more appropriate.

We now provide several evidences supporting our de Sitter swampland conjecture \eqref{eqn:conjecture-DBI}-\eqref{eqn:def-Omega} for the single-field DBI model.

First, instead of the Lagrangian in \eqref{eqn:action-twofields}, we could also start with more general forms of the Lagrangian of two-field models. For instance, we can multiply the kinetic term for $\chi$ with a function of $\chi$. In the large $\beta^2$ limit, the equation of motion for $\chi$ will again lead to the attractor solution in \eqref{eqn:eom-chi2}, and thus the following derivations in section \ref{secHS} and \ref{secDBI} are still valid.

Second, as already mentioned at the end of section \ref{secPX}, the prescription for a more general single-field model with the Lagrangian $P(X,\varphi)$ correctly recovers our de Sitter conjecture for the single-field DBI model. This can be considered as a rather non-trivial supporting evidence for our conjecture,  considering the fact that the two-field model considered in section \ref{secPX}, when specialized to $P(X,\varphi)=T(\varphi)\left[ -\sqrt{1-\frac{2X}{T(\varphi)}} + 1\right] - U(\varphi)$, is not the same as the one considered in section \ref{secHS}. Before taking the limits $Z\to 0$ and $\beta^2\to\infty$, even the field space metrics are different. The two different prescriptions nonetheless give essentially the same de Sitter swampland conjecture for the single-field DBI model.

Third, our de Sitter conjecture reflects the symmetry of the single-field model when available, even if the two-field model does not respect the symmetry. The DBI Lagrangian itself has a nonlinear realization of the 5-dimensional AdS symmetry for a special choice of $T(\varphi)$. On setting $T(\varphi)=\varphi^4/\lambda$, one remnant symmetry in the action is the scaling $\varphi \to b \varphi$, $x^\mu \to x^\mu/b$. It is known that the standard formulas of the scalar and tensor spectral tilts, $n_s$ and $n_t$ (see e.g. \cite{Kobayashi:2007hm}), are invariant under this scaling, provided that not only $\varphi$ and $x^{\mu}$ but also the Planck scale scales as $M_{\rm Pl}\to b M_{\rm Pl}$. We can see that our conjecture \eqref{eqn:conjecture-DBI}-\eqref{eqn:def-Omega} is invariant under this scaling.

In the present paper we have used a two-field model as a bridge between the de Sitter swampland conjecture and the single-field DBI model. In particular, we have shown that in the large $\beta^2$ limit the second field becomes infinitely heavy and thus can be integrated out. This means that in the regime of validity of the effective field theory, the two-field system is equivalent to the single field DBI scalar system both at the classical and quantum level.
In particular, they should give the same predictions for all correlation functions~\footnote{There is subtlety about at which level $\chi$ should be heavy for the validity of the single field effective theory. Roughly speaking, if the inverse of the mass of $\chi$ is smaller than the time scale of the process that excites $\chi$, the single field effective theory is valid. (see e.g. \cite{Gao:2012uq}). Since we take the limit where $\chi$ is infinitely heavy, the single field effective theory is valid all the way up to the cutoff scale of the two-field model.}.

Some of our intermediate results are helpful not only for the extension of the de Sitter swampland conjecture to the models with non-canonical kinetic terms, but also for the extension to the multi-field models with linear kinetic terms. Especially, we found that if the field space is highly curved, the leading contribution to the lowest mass squared does not agree with the lowest eigenvalue of the Hessian of the potential, as the field space curvature also contributes to the mass term. Together with the fact that the existence of the friction term changes the stability of the scalar perturbations, we propose that the last inequality in the refined de Sitter swampland conjecture should involve the lowest mass squared for scalar perturbations that can be also significantly different from the lowest eigenvalue of the Hessian of the potential. 
 
Finally, it is interesting to note that the first requirement in (\ref{eqn:conjecture-kessence}) is much relaxed for models with smaller values of $P_{,X}$, satisfying $|P_{,X}| \ll P_{,XX}X$, such as ghost condensation/inflation~\cite{ArkaniHamed:2003uy,ArkaniHamed:2003uz,Senatore:2004rj}~\footnote{We thank Justin Khoury for pointing this out.} (see \cite{Mukohyama:2009rk,Mukohyama:2009um,Jazayeri:2016jav} for issues related to ghost condensation/inflation and the swampland). As we have already discussed at the end of Section \ref{secSW}, the values of $c_{0,1,2}$ have some uncertainty and could be very small in concrete low energy realizations of quantum gravity theories, and thus it is still premature to claim any contradictions with the inflation and dark energy models. Nonetheless, if $c_{0,1,2}$ are finally proved to be of $\mathcal{O}(1)$, then our discussion on the nonlinear kinetic terms provides a plausible direction of constructing the inflation and dark energy models compatible with the conjecture and observations. Also, it is worthwhile to consider further extensions of the conjecture to more general theories, for instance, the Galileon models~\cite{Nicolis:2008in}, which we will leave for future work.

%%%%%%%%%%%%%%%%%%%%%%%%%%%%%%%%%%%%%%%%%%%%%%%%%%%%%%%%%%%%%%%%%%%%%%%%%%%%%%%%%%%%%%%%%%%%%%%%%%%%%%%%%%%%%%%%%%%%%%%%%%%
\acknowledgments

The authors thank participants of Two-day Focus Meeting on ``Quantum Entanglement in Cosmology'' (KAKENHI Grant Nos. 15H05888 and 15H05895) held at Kavli IPMU for useful comments. S. Mizuno and SP are grateful to the generous hospitality of YITP, Kyoto University during this collaboration. 
SP and YLZ thank the hospitality of School of Physics and Technology, Wuhan University during their visit.
 The work of S. Mukohyama was supported by Japan Society for the Promotion of Science (JSPS) Grants-in-Aid for Scientific Research (KAKENHI) No. 17H02890, No. 17H06359. 
SP was supported by the MEXT/JSPS KAKENHI No. 15H05888.  S. Mukohyama and SP were also partially supported by the World Premier International Research Center Initiative (WPI Initiative), MEXT, Japan.
YLZ was supported by Grant-in-Aid for JSPS international research fellow(18F18315).

\appendix

\section{Squared masses from homogeneous perturbations}
\label{AppA}
In the decoupling limit $M_{\rm Pl}\to \infty$, one can simply study the action (\ref{eqn:action-twofields}) for the two scalar fields in the fixed Minkowski spacetime without coupling the system to gravity. Introducing homogeneous perturbations around a homogeneous background as
\begin{eqnarray}
 \varphi & = & \varphi^{(0)}(t) + \delta\varphi(t)\,,\nonumber\\
 \chi & = & \chi^{(0)}(t) + \delta\chi(t)\,,
\end{eqnarray}
expanding the action up to second order in perturbations, performing some integrations by part, and using the attractor background (\ref{eqn:background-attractor}), we find the quadratic action of the following form in terms of $\mathcal{K}$, $\mathcal{M}$ and $\mathcal{V}$ defined in (\ref{eqn:def-MKV}). 
\begin{equation}
 I^{(2)} = \frac{1}{2}\int dt \left[ \dot{Y}^{\rm T}\mathcal{K}\dot{Y} + \dot{Y}^{\rm T}\mathcal{M}Y +Y^{\rm T}\mathcal{M}^\text{T}\dot{Y} - Y^{\rm T} \tilde{\mathcal{V}} Y\right]\,, \quad Y = \left(\begin{array}{c}
					    \delta\varphi \\
																								    \delta\chi
																								   \end{array}\right)\,,
\end{equation}
where $\tilde{\mathcal{V}} = \lim_{M_{\rm Pl}^2\to\infty}\mathcal{V}$. Therefore the squared masses for the homogeneous perturbations are determined by 
\begin{equation}
 \det \left[ m^2 \mathcal{K} - 2im\mathcal{M} - \tilde{\mathcal{V}} \right] = 0\,.
\end{equation}
The solutions to this equation agree with (\ref{eqn:mpm2})-(\ref{eqn:Omega}) in the decoupling limit $M_{\rm Pl}^2\to \infty$.

\end{document}